\documentclass[superscriptaddress,nofootinbib,preprint]{revtex4}
\usepackage{graphicx}
\usepackage{amssymb,amsmath,amsfonts}
\usepackage[hypertex]{hyperref}

\def\lozenge{\boxit{\hbox to 1.5pt{\vrule height 1pt width 0pt \hfill}}}

\def\veps{\varepsilon}
\def\eg{{\it e.g.}}
\def\etc{{\it etc}}

\def\mpl{\ifmmode M_{pl}\else $M_{pl}$\fi}
\def\mpl{\ifmmode \overline M_{Pl}\else $\bar M_{Pl}$\fi}

\newcommand{\ord}[1]{{\cal O}( {#1})}

\newcommand{\gsim}{\gtrsim}

\newcommand{\sz}{S^1/Z_2}

\newcommand{\dalam}{\raise-1mm\hbox{\large$\Box$}}

\newcommand{\beq}{\begin{equation}}
\newcommand{\eeq}{\end{equation}}

\begin{document}

% page numbers bottom-center
\pagestyle{plain}

\hfill$
\vcenter{ 
%\hbox{\bf BNL xxxx}
\hbox{\bf SLAC-PUB-13411}} $

\vskip 1cm
\title{New Dimensions for Randall-Sundrum Phenomenology}

\author{Hooman Davoudiasl}
\email{hooman@bnl.gov} \affiliation{Department of Physics,
Brookhaven National Laboratory, Upton, NY 11973-5000, USA}

\author{Thomas G. Rizzo\footnote{Work supported in part
by the Department of Energy, Contract DE-AC02-76SF00515.}}
\email{rizzo@slac.stanford.edu}
\affiliation{Stanford Linear Accelerator Center, 2575 Sand Hill Rd.,
Menlo Park, CA  94025, USA}
%%%%%%%%%%%%%%%%%%%%%%%%%%%%%%%%%%%%%%%%%%%%%%

\begin{abstract}

We consider a 6D extension of the Randall-Sundrum (RS) model, RS6, where
the Standard Model (SM) gauge fields are allowed to propagate in an
additional dimension, compactified on $S^1$ or $\sz$.  In a minimal
scenario, fermions propagate in the 5D RS subspace and their
localization provides a model of flavor.  New Kaluza-Klein (KK)
states, corresponding to excitations of the gauge fields 
along the 6$^{\rm th}$
dimension, appear near the TeV scale.  The new
gauge KK modes behave differently from those 
in the 5D warped models.  These RS6 states 
have couplings with strong dependence on 
5D field localization and, within the SM, only interact  
with heavy fermions and the Higgs sector,  
to a very good approximation.  
Thus, the collider phenomenology of
the new gauge KK states sensitively depends on the 5D fermion
geography. We briefly discuss inclusion of SM fermions in all 6
dimensions, as well as the possibility of going beyond 6D.

\end{abstract}
\maketitle

%%%%%%%%%%%%%%%%%%%%%%%%%%%%%%%%%%%%%%%%%%%%%%%%%%%%%%%%%%%%%%%%%%%%%%%%%%%

\section{Introduction}

The Randall-Sundrum (RS) model \cite{Randall:1999ee} has been
extensively discussed as a resolution of the hierarchy between the
$\ord{\rm TeV}$ weak scale in the Standard Model (SM) and the scale
$M_P\sim 10^{19}$~GeV of 4D gravity.   The original model
\cite{Randall:1999ee} was based on a slice of AdS$_5$, bounded by
two 4D Minkowski branes.  This model only addressed the weak-$M_P$
hierarchy, using the exponentially warped 5D spacetime, with 4D SM
fields localized at the TeV (IR) brane and the 4D gravity localized
near the Planck (UV)  brane. It was later shown that extending the
SM content to all 5 dimensions
\cite{Davoudiasl:1999tf,Pomarol:1999ad,Grossman:1999ra,Gherghetta:2000qt}
still allows one to address the hierarchy, as long as the Higgs
sector is localized near the TeV brane.  An interesting consequence
of this extension is that 5D fermion masses allow one to localize
the zero modes of these fields along the 5$^{\rm th}$ dimension
\cite{Grossman:1999ra} and provide a predictive model of flavor
\cite{Grossman:1999ra,Gherghetta:2000qt}.

Various extensions of the RS model have been considered in the
literature.  Much of the discussion has been concerned with
expanding the bulk field content and extensions of 5D gauged
symmetries, in order to enhance the agreement of the of the model
with low energy data. In comparison, less attention has been devoted
to extending the geometrical basis of warped models.

The RS model can be considered to be an effective theory, emerging  from a
string theoretic construction.  Also, the AdS/CFT correspondence 
\cite{Maldacena:1997re} has
been very helpful in relating the geometric results in the RS
picture to those arising from 4D strong dynamics 
\cite{ArkaniHamed:2000ds,Rattazzi:2000hs}.  These theoretical
aspects are generally contained in a larger picture with more than 5
dimensions.  Viewed in this way, it is natural to consider adding
additional dimensions to the RS geometry and studying their potential
observable consequences. Some work along this direction can be found
in Refs.~\cite{Kogan:2001yr,Multamaki:2002cq,Davoudiasl:2002wz,Bao:2005ni},
where dimensions beyond the original 5D have been considered.
However, these works, by and large, have concentrated on the gravitational
sector and its phenomenology.

In this paper, we consider extending RS-type models with additional
non-warped dimensions, where the gauge sector of the SM is also
allowed to propagate in all dimensions.  We adopt a minimal setup,
where SM fermions are only allowed to travel in the original warped
5D spacetime, in order to address flavor physics, but some or all
gauge fields are allowed to reside in 6 dimensions.  We will
consider an $S^1$ or $\sz$ compactification for the 6$^{\rm th}$
dimension and refer to this extended geometry as RS6. We concentrate
on a 6 dimensional $SU(3)_c$ sector and show that new Kaluza-Klein
(KK) fields emerge at the TeV scale, with couplings very different
from those that arise in the usual 5D picture 
\cite{Gherghetta:2000qt,Davoudiasl:2000wi}.  The couplings of these new
modes to 5D fermion fields are shown to be exponentially sensitive to
localization along the 5$^{\rm th}$ dimension.   We then begin to consider the 
collider phenomenology of these new KK modes at the LHC and their
potential for discovery.

In the next section, we derive the KK equation of motion and the
spectrum, for a gauge field in RS6 .  We also briefly
discuss extensions to higher dimensional spheres. In section III, we
derive the couplings of these KK modes to 5D fermions.  In section IV, we discuss
the LHC phenomenology of this model and outline its discovery
prospects. We will also briefly discuss possible extensions of our
minimal RS6 model to RS$n$, $n>6$, as well as scenarios with
fermions in more than 5 dimension.  We will conclude in section V.
The appendix provides some relevant expressions for the RS7 case
with an $S^2$ compactification.

\section{KK Spectrum and Wavefunctions}

Lets us consider the RS6 metric $G_{MN}$,
$M,N=0,1,\ldots,5$; $x_4 = r_c \phi, x_5 = R\theta$,
with an extra dimension compactified on $S^1$:
\beq
ds^2 = e^{-2\sigma}\eta_{\mu \nu}  dx^\mu dx^\nu - r_c^2 d\phi^2 - R^2 d\theta^2,
\label{metric}
\eeq
where, as usual $\sigma(\phi) = k r_c |\phi|$, $k$ is the scale of curvature and $r_c$ is the radius
of compactification of the AdS$_5$ slice; $\phi \in [-\pi, \pi]$ and a $Z_2$ orbifolded
5$^{\rm th}$ dimension is assumed.  Here, $R$ is the radius of
$S^1$ and $\theta \in [0, 2\pi]$; in the absence of fine-tuning, it is natural to imagine that, \eg, $kR\sim 1$.  
The choice of the 6D energy momentum tensor consistent
with this background has been discussed in Ref.~\cite{Davoudiasl:2002wz}, 
where the corresponding gravitational sector was studied.

The action for a 6D non-interacting gauge field is given by
\beq
S_A = -\frac{1}{4} \int  r_c d \phi \int R d\theta \, \sqrt{ -G}\, G^{AM}G^{BN} F_{AB} F_{MN},
\label{SA}
\eeq
where $G = \det(G_{MA})$ and $F_{MN} = \partial_M A_N - \partial_N A_M$.
As is well-known, with 2 extra dimensions, in addition to the 4D gauge fields 
there is also a 4D tower of KK scalars that correspond to
a combination of $A_\phi$ and $A_\theta$.  For example, in the case of $SU(N)$, this would
correspond to a tower of massive adjoint scalars without a zero-mode.  
As is also well-known in the case of flat extra dimensions, 
such scalars can lead to their own interesting new physics{\cite {UED6}}. 
Here, we will concentrate on the 4D vector modes $A_\mu$,
$\mu=0, 1, 2, 3$,  and set $A_\phi = A_\theta = 0$.  With this choice, the action (\ref{SA}) yields
\beq
S_A = \int  r_c d \phi \int R d\theta \left\{ -\frac{1}{4}  F^{\mu \nu}F_{\mu\nu} - \frac{1}{2}
\left[\frac{1}{r_c^2}\partial_\phi \left(e^{-2\sigma}\partial_\phi A^\mu\right)A_\mu
+ e^{-2\sigma}\frac{1}{R^2} \partial_\theta^2 A^\mu A_\mu \right]\right\}.
\label{SAmu}
\eeq

The vector field $A_\mu(x, \phi, \theta)$ can be expanded in KK modes
\beq
A_\mu(x, \phi, \theta) = \sum_{n, l}
A^{(n, \, l)}_\mu(x)
\frac{\chi^{(n,\, l)}(\phi)}{\sqrt {r_c}}
\frac{\varphi^{(l)}(\theta)}{\sqrt {R}}.
\label{AKK}
\eeq
The $\theta-$dependent wavefunction is given by
\beq
\varphi^{(l)}(\theta)= e^{i l \theta}/{\sqrt {2 \pi}}
\eeq
in the case of $S^1$ and
\beq
\varphi^{(l)}(\theta)= \left\{ \begin{array}{ll}
                               {1/\sqrt{2\pi}}, & \quad l=0\\
                               {\cos (l\theta)/\sqrt\pi}, & \quad l\ne 0
                              \end{array}
                       \right.
\eeq
for the orbifolded $\sz$ case.
The wavefunctions obey the orthonormality conditions
\beq
\int d\phi \,\chi^{(m,\, l)}\chi^{(n,\, l)} = \delta_{m n}
\label{orthophi}
\eeq
and
\beq
\int d\theta \,\varphi^{(l)}(\theta)\varphi^{(l^\prime)}(\theta)
= \delta_{l l^\prime}.
\label{orthotheta}
\eeq
Inserting the above KK expansion into the action (\ref{SAmu}),
we find the following eigenvalue equation for the $(n,l)$ mode
of mass $m_{nl}$
\beq
-\frac{1}{r_c^2} \frac{d}{d\phi}\left(e^{-2 \sigma}
\frac{d}{d\phi} \chi^{(n,\, l)}(\phi)\right) +
e^{-2 \sigma}\left(\frac{l}{R}\right)^2 \chi^{(n,\, l)}(\phi)
= m_{nl}^2 \,\chi^{(n,\, l)}(\phi).
\label{KKeq}
\eeq

The above equation of motion corresponds to that
for a 5D vector field of bulk mass $l/R$, in the RS background.
The solutions are given by \cite{Pomarol:1999ad}
\beq
\chi^{(n,\, l)}(\phi) = \frac{e^\sigma}{N_{nl}} \,
[J_\nu(z_{nl}) + \alpha_{nl} Y_\nu(z_{nl})],
\label{chi}
\eeq
where $J_\nu$ and $Y_\nu$ denote Bessel functions of order $\nu$ where 
\beq
\nu \equiv \sqrt {1 + \left(\frac{l}{k R}\right)^2},
\label{nu}
\eeq
with $z_{nl}(\phi) \equiv (m_{nl}/k) e^\sigma$.
We will define for simplicity the combination 
\beq
\zeta_\nu (z_{nl}) \equiv J_\nu(z_{nl}) + \alpha_{nl} Y_\nu(z_{nl}).
\label{zeta}
\eeq
We then impose the boundary conditions $\partial_\phi \chi^{(n,\, l)}(\phi)  = 0$ at
$\phi = 0, \pi$, which yield
\beq
z_{nl} \zeta_\nu^\prime + \zeta_\nu = 0.
\label{bc}
\eeq
Using Eq.~(\ref{orthophi}) and Eq.~(\ref{bc}), we find for the normalization
\beq
N_{nl} = \frac{e^{kr_c\pi}}{x_{nl} \sqrt{k r_c}}
\sqrt{\zeta_\nu \left[z_{nl}^2 - (\nu^2-1)\right]\left|^{x_{nl}}_{\veps_{nl}}\right.},
\label{Nnl}
\eeq
where $x_{nl} = z_{nl}(\pi)$ and $\veps_{nl} = z_{nl}(0)$.

Eq.~(\ref{bc}) evaluated at $\phi=0$ can be used to determine the coefficients 
$\alpha_{nl}$: 
\beq \alpha_{nl} = - \frac{J_\nu (\veps_{nl}) +
\veps_{nl} J^\prime_\nu (\veps_{nl})} {Y_\nu (\veps_{nl}) +
\veps_{nl} Y^\prime_\nu (\veps_{nl})}. 
\label{alpha} 
\eeq 
One can then easily show that $\alpha_{nl} \sim \veps_{nl}^{2\nu}$.  Since
addressing the hierarchy implies $\veps_{nl} \sim 10^{-16}$, for
$\nu > 1$, one can safely ignore the part of the wavefunction
$\chi^{(n,\, l)}$ that is proportional to $\alpha_{nl}$.  The masses
of the KK modes corresponding to $\chi^{(n,\, l)}$ are given by
$m_{nl} = x_{nl}\, k\, e^{-k r_c \pi}$, where $x_{nl}$ are the roots
of the transcendental equation \beq J_\nu (x_{nl}) + x_{nl} J^\prime_\nu (x_{nl}) = 0,
\label{roots} \eeq obtained from Eq.~(\ref{bc}) at $\phi = \pi$,
ignoring terms proportional to $\alpha_{nl}$.   In this
approximation, we then find
\beq
\chi^{(n,\, l)}(\phi) \simeq
\frac{e^\sigma}{N_{nl}} \, J_\nu(z_{nl}),
\label{chiapprox}
\eeq
where
\beq N_{nl} \simeq \frac{e^{kr_c\pi}}{\sqrt{k r_c}}\,
\beta(x_{nl}, l)  J_\nu(x_{nl}) ,
\label{Nnlapprox}
\eeq and
\beq
\beta(x_{nl}, l) \equiv
\left[1- \left(\frac{l}{k R
x_{nl}}\right)^2\right]^{1/2}.
\label{beta}
\eeq
For the purposes of this work, Eqs.~(\ref{roots}),
(\ref{chiapprox}), and (\ref{Nnlapprox}) are very good
approximations and will be used in what follows. Note that since $\beta$ is a 
real quantity we must have $x_{nl}>l/kR$ and thus states with $l>0$ do not have 
zero-modes. 

One can similarly derive expressions for KK gauge fields from
compactification on $S^N$, with $N>1$.  As an example, we display 
the wavefunctions for the case of $S^2$ in the appendix. Much of what
follows in the next section can then be applied to higher dimensional spherical compactifications, 
with rather straightforward modifications.

\section{KK Couplings in the Minimal RS6 Model}

Here, we consider a minimal extension where gauge fields are allowed to propagate
in all 6 dimensions and the SM fermions reside in the 5D RS subspace, in order to explain
the flavor structure observed at low energies.  Later we will discuss typical fermion ``geographies"
and their experimental consequences in the context of this minimal RS6 model.
For simplicity we will assume that the SM fermions are localized at
$\theta = 0$. To get some sense of the
magnitude of the couplings to SM fermions, we will first 
consider two extreme cases, with a fermion localized at either the 
UV or the IR brane.  At the UV brane, $\phi=0$, and we have (for $l>0$) 
\beq
\chi^{(n,\, l)}(0) \sim \veps_{nl}^{\nu+1},
\label{chi0}
\eeq
whereas at the IR brane, $\phi=\pi$, we have for all $l$ 
\beq
\chi^{(n,\, l)}(\pi) \simeq \sqrt{k r_c} /\beta(x_{nl}, l).
\label{chipi}
\eeq

\begin{figure}[htbp]
\centerline{
\includegraphics[width=7.5cm,angle=90]{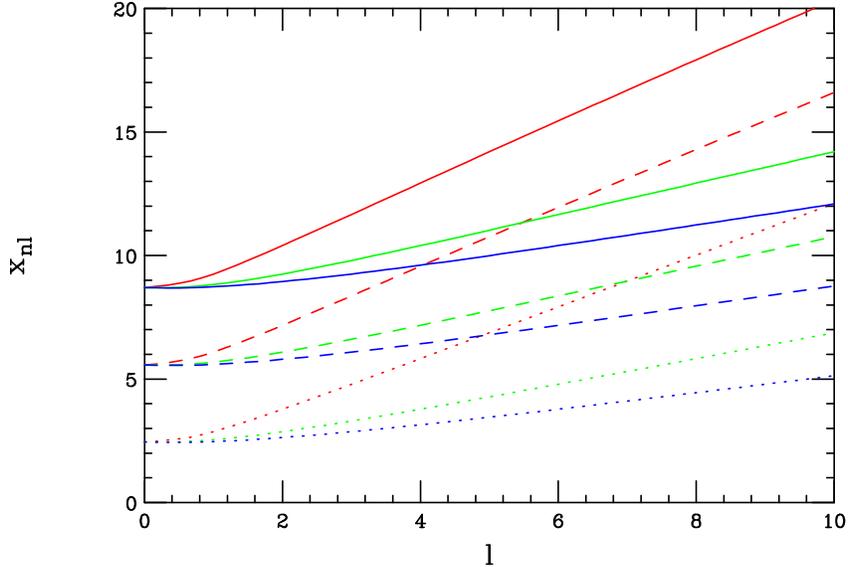}}
\vspace*{0.1cm}
\caption{The lowest lying roots, $x_{nl}$, assuming $kR=1$(red), 2(green) or 3(blue) for $n=1$(dots), 2(dashes) 
or 3(solid) as a function of $l$.}
\label{fig0}
\end{figure}

Using the zero mode wavefunction $\chi^{(0,\, 0)} = 1/(2 \pi)$, one
can easily derive the relation \beq g_4 = \frac{g_6}{2 \pi \sqrt{r_c
R} }\, , \label{g4} \eeq between the 4D and 6D gauge couplings,
$g_4$ and $g_6$, respectively.  Using Eq.~(\ref{chi0}), one then
finds that the $l\neq 0$ KK modes {\it exponentially decouple} from
fermions at the UV brane.  However,  the couplings $g_{nl}$, for
$l\neq 0$, to the IR brane fermions are given by \beq g_{nl}|_{\rm
IR}= \left\{ \begin{array}{ll}
                               {g_4 \sqrt{2 \pi k r_c}/\beta(x_{nl}, l)}, & \quad S^1\\
                               {g_4 \sqrt{4 \pi k r_c}/\beta(x_{nl}, l)}, & \quad \sz\, ,
                              \end{array}
                       \right.
\label{gnl}
\eeq
where we have used Eq.~(\ref{chipi}).
Given that $1/\beta(x_{nl},l)>1$ and $k r_c \sim 10$, we see that $g_{nl}\gsim 8 g_4$ at the IR brane is quite large.
Note that this coupling is larger by factors of $1/\beta(x_{nl},l)$, for $S^1$, and
$\sqrt{2}/\beta(x_{nl},l)$, for $\sz$, than the corresponding coupling in the 5D RS model.  For
example, if we take $kR = 1$,  we obtain $x_{10}\simeq 2.45$, $x_{11}\simeq 2.87$, and $1/\beta(x_{11}, 1) \simeq 1.07$. 
Hence, the ratio of the mass of the lightest $l\neq 0$ KK mode to the lightest KK mode is $2.87/2.45 \simeq 1.17$ in 
this case.  
Fig.~\ref{fig0} shows some of the lowest lying roots obtained from Eq. 16, which determines the gauge KK masses,  
as a function of the value of $l$. These will be important when we study the possible LHC signatures in the next Section. 

Having studied the extreme UV/IR brane cases, let us consider the intermediate cases where the fermions
are not confined to either brane, but have 5D profiles.  For a zero mode fermion, 
the bulk profile is given by \cite{Grossman:1999ra,Gherghetta:2000qt}
\beq
f_0 = \frac{e^{- c \sigma}}{N_0},
\label{f0}
\eeq
with the normalization
\beq
N_0 = \left[\frac{e^{k r_c \pi(1- 2c)} - 1}{k r_c(1/2- c)}\right]^{1/2}.
\label{N0}
\eeq
Here, $c>1/2$ corresponds to UV localization (light fermions) and $c<1/2$ corresponds
to IR localization (heavy fermions).

The coupling $g_{nl; c}$ of the
$(n,l)$ modes to a 5D zero mode fermion is then given by
\beq
g_{nl; c} = \sqrt{2 \pi} \int d\phi f_0^2  \, \chi^{(n,\, l)},
\label{gnlc}
\eeq
for $S^1$, and the corresponding $\sz$ value is larger by $\sqrt{2}$ for $l\ne 0$. 
For example, if we choose typical values $c=0.6$ for light fermions and
$c=0.3$ for heavy fermions and set $k R = 1$,  we then find $g_{11;  0.6} = 9\times 10^{-4} g_4$ and
$g_{11; 0.3} = 2.1 g_4$, for the $S^1$ compactification.  Thus, typically, we expect the $l\neq 0$ modes to decouple 
from light fermions, to a good approximation.  However, the coupling of these modes to light
fermions has a strong dependence on the exact fermion localization in 5D.
{\footnote {Note that this may lead to additional 
flavor issues due to the exchanges of these states, but such a discussion 
is beyond the scope of the present paper.}}
This is in contrast to the original RS
model, corresponding to $l=0$ here,
where gauge KK couplings to light fermions are nearly universal and small, but not negligible; for example, 
taking $c=0.6$ one obtains instead $g_{10; 0.6} = 0.19 g_4$.

\section{LHC Phenomenology}

Here we address potential signals of the RS6 model at the LHC.  Similar considerations can be applied
to higher dimensional spherical compactifications $S^N$, with $N>1$, keeping the SM fermions in the 5D subspace.
For the 5D RS model, corresponding to $l=0$ modes,  there is very little sensitivity
in the gauge KK couplings to the UV localization ($c>1/2$) of light fermions and one can choose a typical
value for $c\gsim 1/2$ and obtain the universal KK coupling to light fermions of the SM.  This allows
for a relatively model independent assessment  of the relevant collider phenomenology
\cite{Agashe:2006hk,Lillie:2007yh,Agashe:2007ki}.  However, as we saw before, the
couplings of the $l\neq 0$ gauge KK modes are very sensitive to the 5D localization of
the fermions.  Thus, to study possible signatures of the RS6 model we must be more specific about
the localization parameters for the important initial state fermions at colliders.
In what follows we consider the simplest case of 
$S^1$ compactification.

\begin{figure}[htbp]
\centerline{
\includegraphics[width=7.5cm,angle=90]{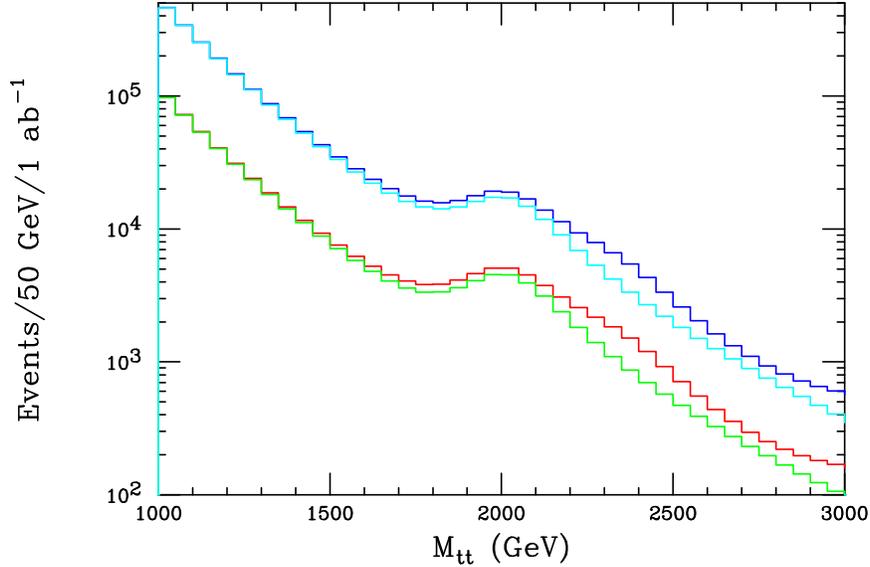}}
\vspace*{0.1cm}
\caption{RS6 KK production for $kR=1$, where the lightest state is at 
$m_{10} = 2$~TeV.  The upper and lower pairs of histograms correspond to cuts of $|\eta| < 1, 1/2$, 
respectively, on the final state tops. Both final state top quarks are also required to have $p_T>200$ GeV. 
In each pair, the upper histogram includes RS6 KK contributions up 
to states which are $\sim 1.54$ times more massive than the lightest KK mode.  The lower histogram in the pair 
represents the usual 5D RS scenario.  An integrated luminosity of $L = 1$~ab$^{-1}$ has been assumed.}
\label{fig1}
\end{figure}

To make a comparison of our results with some of the existing literature easier, we
adopt a 5D flavor model in which $t_R$ is the most IR-localized SM fermion and
couples to the lightest KK mode ($l = 0$) with the strength $g_4 \sqrt{k r_c \pi}$.  Here,
we will concentrate on KK gluons and hence $g_4=g_s$, where $g_s$ is the
$SU(3)$ coupling in the SM.  This corresponds to $c(t_R)= -0.6$, in our convention.  
We choose the localization parameters 
close to those in realistic models \cite{Huber:2003tu}, 
but we only attempt to capture the essential 
features of 4D flavor and not the details.  To get
the correct top mass, we then require $c(Q^3_L)=0.3$, where $Q^3_L$
denotes the third family quark doublet and we have assumed that the
Higgs is on the IR brane.  This way, the dominant decay mode of the new states is
into the channel $t_R{\bar t_R}$.  However, in the following, we would like to address resonant
production of the $l\neq 0$ modes from $q {\bar q}$ initial states. For very light quarks we saw 
above that these couplings were very small and so, \eg, conventional $u\bar u, d\bar d$ and $s\bar s$ initial  
state partons lead to small cross sections. Given that $b_L$
is IR localized with our choice of parameters, its coupling to the $(11)$ state
is $\sim 2 g_s$ and fairly large.  Hence, it makes sense to examine
whether one can use the $b$-content of the initial states for KK production.  Here,
we ignore higher order corrections to the flux of initial state $b$-quarks, which is
a good approximation for our purposes \cite{Campbell:2004pu}.

Ignoring the rest of the quarks, for $kR=1$, we found, however, 
that the small $b$-content of the initial state
protons does not yield a significant signal for the $l\neq 0$ modes, with an integrated 
luminosity $L = 1$~ab$^{-1}$.  However, we
have determined that the inclusion of the charm content, together with that of bottoms, 
of the proton plays an important role here.
Choosing $c(c_R)=0.52$ for the singlet charm quark $c_R$, we find that its coupling to
the lightest $l\neq 0$ states is roughly $0.07 g_s$.  Even though this is not very large, it turns
out that the much larger charm content of the proton compensates for it with the enhanced top quark coupling 
in the final state.  In Fig.~(\ref{fig1}),
we have presented the result for the case $kR=1$, choosing the lightest state $(10)$ to be
at 2~TeV, and including the effects of coupling to $c_R$ in the proton; here and for the 
rest of this discussion $L = 1$~ab$^{-1}$ is assumed. Given the couplings above, very roughly, all of the KK 
states have width-to-mass ratios of $\simeq 1/6$. Note in addition that for the $S^1$
compactification there are two degenerate states at each $l\neq 0$ which is a significant source of cross section 
enhancement.  The upper pair
of curves correspond to the pseudo-rapidity cut $|\eta|<1$ and the lower one corresponds
to $|\eta|<1/2$ on the top quarks in the final state.  
In each pair, the upper histogram includes the contributions of $l\neq0$ resonances, up to
a state that is $3.78/2.45\simeq 1.54$ times heavier than the $(10)$ KK mode.  The lower histogram in
the pair does not include any contribution from $l\neq0$ modes and corresponds to the
usual 5D RS result.  We see from the figure that when $kR=1$ there is no obviously 
clear signal for RS6 versus the usual RS.  Note that we have not included here the effects of boosted top jets, 
branching fractions, 
efficiencies and detector resolutions that go into the actual extraction of the signal from
the data.  As these effects will make it only more difficult to distinguish the two cases, we conclude
that the signal in this case is at best only very marginal.

\begin{figure}[htbp]
\centerline{
\includegraphics[width=7.5cm,angle=90]{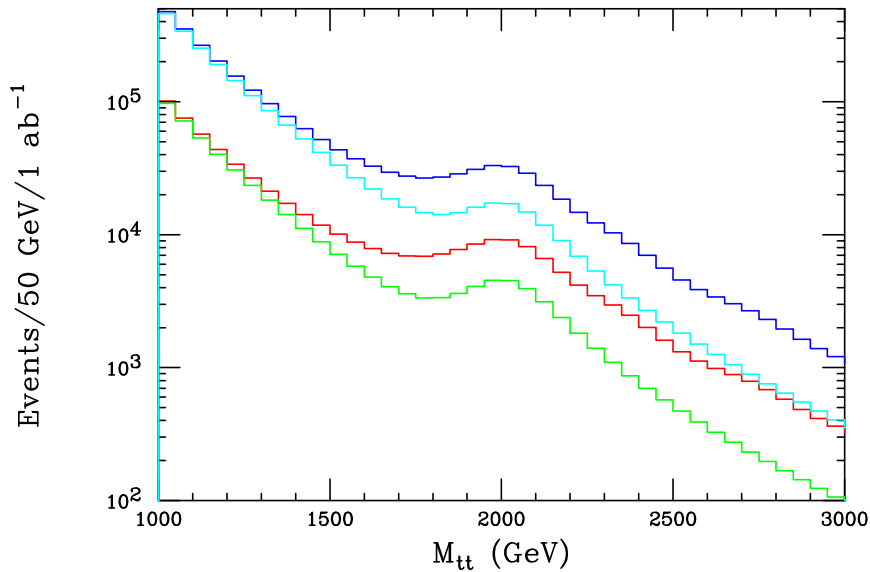}}
\vspace*{0.1cm}
\caption{Same as the previous figure but now with $kR=2$.}
\label{fig2}
\end{figure}

\begin{figure}[htbp]
\centerline{
\includegraphics[width=7.5cm,angle=90]{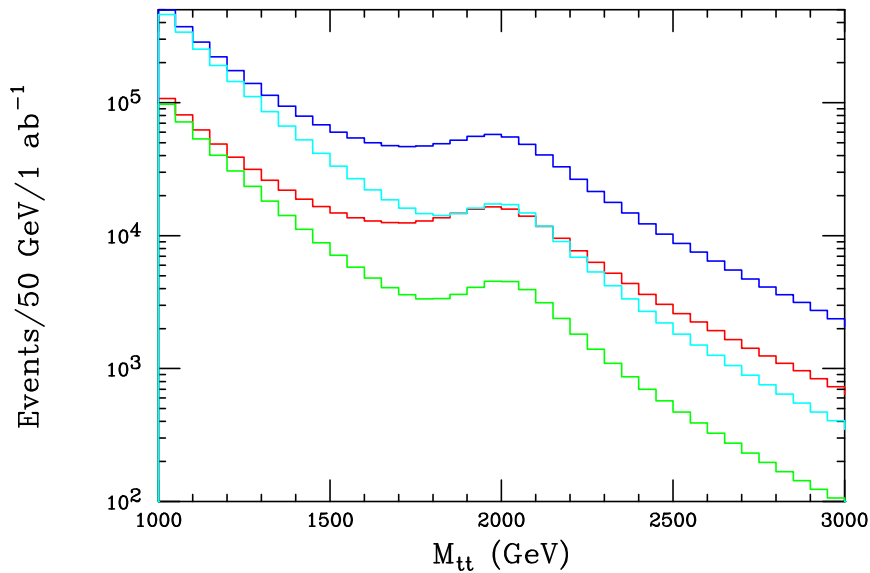}}
\vspace*{0.1cm}
\caption{Same as the previous figure but now with $kR=3$.}
\label{fig3}
\end{figure}

One of the factors that made the signal in the previous case, with $kR=1$,
difficult to observe was that only the additional resonances for $l=1,2$ were included
in the mass range above the $(10)$ state.  Going to the case $kR=2$, still a modest value and a reasonable
choice, will increase the number of contributing resonances
in our mass window and will significantly boost the signal.  For  this values of $kR$, the 
modes included in the mass range $2.0-3.1$~TeV, considered 
in Fig.~(\ref{fig1}),  correspond to $l=1,\ldots,4$.  The expected signal in this case is presented
in Fig.~(\ref{fig2}), using the same cuts as before.  We see that the signal is
now much more pronounced and corresponds to a noticeably different
line shape above the $(10)$ resonance; this is due to the overlap of the 
contributions of multiple resonances which are 
each rather wide.  Given the statistics inferred from
the plot, we would expect a clear signal for RS6 in this case even when efficiencies \etc ~are included.   
Fig.~(\ref{fig3}) is the same
as Fig.~(\ref{fig2}), but now for the case $kR=3$.   As expected, the RS6 signal is now significantly
distinct from the RS case, since all states corresponding to $l=, 1, \ldots, 6$
now contribute in the above mentioned mass interval. Clearly as $kR$ increases further 
the deviation from the classic RS signature will only increase. Now that we see the pattern of change 
induced by the $l\ne 0$ states as we vary $kR$, it is clear that for values of $kR <1$ the 
new gauge KK states will be essentially invisible in the $t\bar t$ channel.

\begin{figure}[htbp]
\centerline{
\includegraphics[width=7.5cm,angle=90]{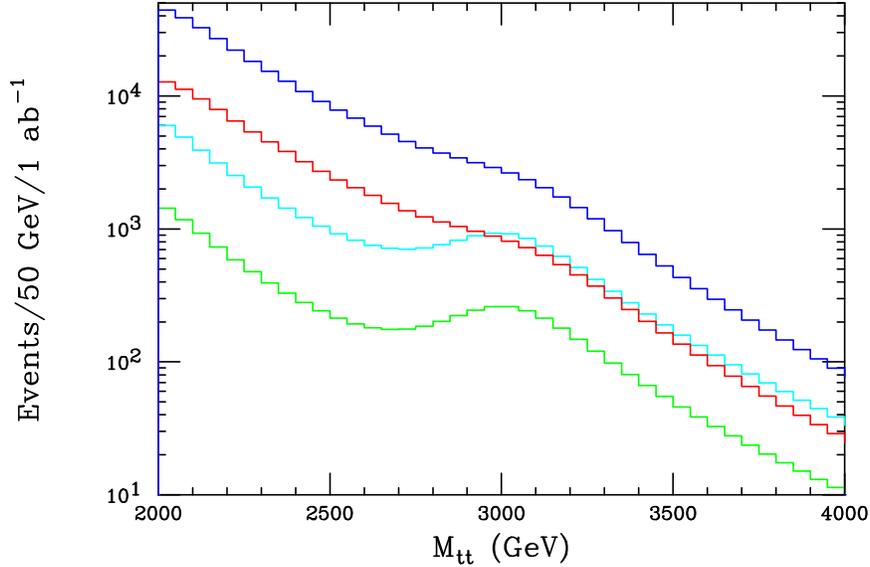}}
\vspace*{0.1cm}
\caption{Same as the previous figure with $kR=3$ and $m_{10}=3$~TeV.}
\label{fig4}
\end{figure}

Here we note that if RS-type models are to emerge at the TeV-scale,
precision data strongly suggest that new symmetries need to be
imposed on these models \cite{Agashe:2003zs,Agashe:2006at}.  Even
then, the mass $m_{10}=2$~TeV of the lightest gluon KK state chosen
for the above plots may not be consistent with current bounds on the
RS model from precision data
\cite{Agashe:2004ay,Agashe:2004cp,Carena:2007ua,Csaki:2008zd,Casagrande:2008hr}.
Hence, in Fig.~(\ref{fig4}), we present the $kR=3$ case with
$m_{10}=3$~TeV, for which the model is in better agreement with the
electroweak precision data (agreement with the flavor data
\cite{Bona:2007vi} typically requires further model building
\cite{Fitzpatrick:2007sa,Csaki:2008eh}). Here, again, the plot
suggests that the signal will be quite prominent and distinct from
the usual RS expectation even after efficiencies \etc ~are included. Note that in 
this case the peaking structure found in the 5D RS case is lost.

At this point we would like to discuss some future directions for going 
beyond the present work.  In terms of collider signatures, 
a potentially interesting production channel may be 
radiation of the $l\ne 0$ gauge KK states off a final state top quark, as the new KK modes couple to 
IR localized fields strongly.  Also, within the setup studied here, for each 6D
gauge group, there is a tower of scalar states in the adjoint
representation.    
A suitable framework for 6D gauge-fixing in
the RS6 model can allow one to identify the combination of the gauge
field polarizations that correspond to this tower of physical
scalars.  We did not address this analysis here, and
confined the scope of our project to the vector modes.

A possible extension of the RS6 model involves the inclusion of the
fermions in all 6 dimensions.  We did not consider this possibility
within our minimal model, where 5D fermions are sufficient to
address flavor physics.  However, inclusion of the fermions in the
6D field content will require elimination of unwanted zero modes
from the 4D effective theory, since 6D fermions come with $\pm$
chiralities, each of which can be decomposed into both left- and
right-handed 4D fermions \cite{Appelquist:2001mj}.

Much of what we studied for the new vector KK modes in this work
will go through with straightforward modifications for
compactification on higher dimensional spheres.  We have provided
the relevant formalism in the case of $S^2$, in the appendix.  We 
note that spheres do not allow
for massless fermion zero modes \cite{Camporesi:1995fb,Abrikosov:2001nj} 
%for states carrying non-trivial `angular momentum' 
and hence the appearance of 5D masses is expected to be a general
consequence of compactification on these manifolds. We speculate
that it may be possible to employ this feature in building warped
models of flavor that use 5D masses for localization of fermions. 

\section{Conclusions}

In this work, we considered extending the RS geometry to RS6 which includes an extra dimension 
compactified on $S^1$ or $\sz$.  This is motivated by a UV completion of the RS model 
within string theory, where additional dimensions are present.  We focused on a minimal model 
with a 6D gauge sector and 5D fermions, localized to explain SM flavor.  
We found the spectrum and wavefunctions of the new gauge KK modes, corresponding to excitations 
along the circle.  These new modes have couplings that are more strongly sensitive to the 5D fermion geography  
than do the usual RS gauge KK modes.  For values of the $S^1$ radius that are somewhat large compared 
to the curvature of the slice of AdS$_5$, there are many new KK modes that are tightly spaced above 
the lightest RS KK mode.  We discussed the potential for observation of these modes at the LHC 
and concluded that for reasonable choice of parameters, the usual RS resonance 
line shapes will be sufficiently modified to distinguish RS6 from the conventional 5D scenario.  Future directions 
including other production channels, KK scalar phenomenology, inclusion of fermions in RS6, and 
higher dimensional compactifications were discussed briefly  in the previous section.

\acknowledgments

We thank Sally Dawson for discussions.  H.D. is supported in part by the DOE grant
DE-AC02-98CH10886.

\appendix
\section{Gauge KK wavefunctions for RS7 with \boldmath$S^2$}

We parameterize the metric as
\begin{equation}
ds^2 = e^{-2\sigma} \eta_{\mu \nu} dx^\mu dx^\nu - r_c^2 \,
d\phi^2 - R^2 (d\theta^2 + \sin^2\theta \, d\omega^2),
\label{s2met}
\end{equation}
where we now have $\theta \in [0, \pi]$ and
$\omega \in [0, 2 \pi]$.

Given the spherical symmetry of the compactification manifold,
we choose the following  KK expansion for the gauge field
\begin{equation}
A_{\mu}(x, \phi, \theta, \omega) = \sum_{n, l, m}
A^{(n, \, l, \, m)}_{\mu}(x)
\frac{\chi^{(n,\, l)}(\phi)}{\sqrt {r_c}}
\frac{Y^m_l(\theta, \omega)}{R},
\label{AKKs2}
\end{equation}
where the $Y^m_l(\theta, \omega)$ are the spherical harmonics.
We now have
\begin{equation}
\nu \equiv \sqrt {1 + \frac{l(l + 1)}{(k R)^2}}\, .
\label{nus2}
\end{equation}
With this change the gauge KK masses as are given in the text above. 

By spherical symmetry we may choose any point on the sphere
to place the 5D fields.  A particularly convenient
choice is $\theta = 0$, for which we have
\begin{equation}
Y^m_l (0, \omega) = \sqrt{\frac{2 l + 1}{4 \pi}}
\delta_{m, 0} \,\,.
\label{ykl}
\end{equation}
This coupling is independent
of $\omega$ and that for any $l$ only allows $m = 0$ states to couple.
For the case when the 5D fields are localized at $\theta=\pi$,
there will be an overall factor $(-1)^l$.

\end{document}